\documentstyle[preprint,revtex]{aps}
\begin{document}
\begin{flushright}
\baselineskip=15pt
UU-HEP-92/9\\
IFFI 92-11\\
hep-th/9210110\\
October 20, 1992
\end{flushright}

\begin{title}
{\bf QUANTUM EINSTEIN-MAXWELL FIELDS:\\
A UNIFIED VIEWPOINT \\ FROM THE LOOP REPRESENTATION}
\end{title}
\author{Rodolfo Gambini}
\begin{instit}
Instituto de F\'{\i}sica, Facultad de Ciencias, \\
Tristan Narvaja 1674, Montevideo, Uruguay
\end{instit}
\author{Jorge Pullin}
\begin{instit}
Department of Physics, University of Utah, Salt Lake City, UT 84112
\end{instit}

\begin{abstract}
We propose a naive unification of Electromagnetism and General
Relativity based on enlarging the gauge group of Ashtekar's new
variables. We construct the connection and loop representations and
analyze the space of states. In the loop representation, the
wavefunctions depend on two loops, each of them carrying information
about both gravitation and electromagnetism. We find that the
Chern-Simons form and the Jones Polynomial play a role in the model.
\end{abstract}

\noindent PACS 4.50.+h, 4.60.+n, 4.20.Fy

\eject

\section{Introduction}

The introduction of the Ashtekar New Variables \cite{As} for the
treatment of canonical gravity has opened new hopes that General
Relativity may be nonperturbatively quantized. In particular the Loop
Representation \cite{RoSm} allows us to construct for the first time
physical states of quantum gravity without recurring to minisuperspace
approximations \cite{BrGaPuprl}. Knot theory and in particular the
Jones Polynomial play a crucial role in the theory \cite{BrGaPunpb}.

In spite of these successes it is still not obvious up to which extent
this program can allow us to understand the physics at the energies of
relevance for quantum gravity. In particular, the main results of the
program at the moment seem quite tailored to vacuum General Relativity
in four space time dimensions. This raises the question of up to what
extent is this program a suitable approach for the incorporation of
other interactions.

In this work we will suggest that the idea of a unified theory
described in terms of Ashtekar's new variables is possible and that
several appealing results of the vacuum theory find very naturally
their counterpart in the unified model. We will show, for instance,
that knot theory still plays a crucial role and that the techniques
used to find states for the theory in the vacuum case are still
applicable. In summary, these ideas for quantizing gravity can lead to
interesting new insights also in the case where matter fields are
present and therefore are well suited for understanding the physics of
particles at the energies of unification and not just pure gravity.

The idea of unifying gravity with other forces enlarging
the group of Ashtekar variables is not new \cite{Peldan}. However the
program outlined in this paper is less ambitious than others. We do not
pretend to recover the same form of the constraints as the vacuum ones
just with an enlarged group. We will see, however, that the
``kinematic'' constraints can actually be rewritten in this way. This
will be enough to find several connections between results of the
vacuum theory and the unified theory. The Hamiltonian constraint will
be quite different and we will briefly outline the consequences of
this. Although we will only give details for the Einstein-Maxwell
case, the same ideas can be straightforwardly generalized to
Einstein-Yang-Mills theories for SU(N).

\section{Einstein-Maxwell theory \\ in terms of a U(2) connection}

We begin with a brief summary of the Einstein-Maxwell theory in terms
of Ashtekar's new variables. We will assume a 3+1 decomposition of
spacetime has been performed, slicing it into spatial three- surfaces
$\Sigma$ on which all variables are defined. The variables for the
gravitational part are a (densitized) triad $\tilde{E}^{a}_{i}$, which
determines the spatial metric by $\tilde{\tilde{q}}^{ab} =
\tilde{E}^{a}_{i} \tilde{E}^{b}_{i}$ and an $SU(2)$ connection
$A_{a}^{i}$ as conjugate momentum. For the Maxwell field the variables
are the electric field $\tilde{\bf e}^{a}$ and the vector potential
${\bf a}_{a}$. The dynamics of the theory is pure constraint, and the
constraints (including a cosmological constant $\Lambda$) are
\cite{AsRoTa},

\begin{equation}
\partial_{a}\tilde{\bf e}^{a}=0\label{gaussmax}
\end{equation}
\begin{equation}
D_{a} \tilde{E}^{a}=0\label{gaussasht}
\end{equation}
\begin{equation}
i \sqrt{2} \tilde{E}^{a}_{i} F_{ab}^{i} -{\textstyle {1 \over 2}}
\tilde{\bf e}^{a} {\bf f}_{ab}=0\label{diffeo}
\end{equation}
\FL
\begin{eqnarray}
&&\epsilon^{ijk}\tilde{E}^{a}_{i}\tilde{E}^{b}_{j} F_{ab}^{k} +\Lambda
\det(E)^{2} -\\
&&-{\textstyle {1 \over 8}} det(E)^{-2}
\{\tilde{E}^{a}_{i}\tilde{E}^{b}_{i} \tilde{E}^{c}_{j}
\tilde{E}^{d}_{j} \eta_{acf} \eta_{bdg} [\tilde{\bf e}^{f} \tilde{\bf
e}^{g} +\tilde{\bf b}^{f} \tilde{\bf b}^{g}]\}=0\label{hamilton}\nonumber
\end{eqnarray}
where $det(E)^{2} = {-1\over 3 \sqrt 2}
{\rlap{\lower1ex\hbox{$\sim$}}\eta{}}_{abc} \epsilon^{ijk}
\tilde{E}^{a}_{i} \tilde{E}^{b}_{j} \tilde{E}^{c}_{k}$

Equation (\ref{gaussmax}) is the Gauss Law of the U(1) symmetry.
Equation (\ref{gaussasht}) is the Gauss Law of Ashtekar's formalism,
stemming from the invariance under triad rotations. Equation
(\ref{diffeo}) is the diffeomorphism constraint and equation
(\ref{hamilton}) is the Hamiltonian constraint. Notice that this
constraint can be made polynomial by multiplying by $det(E)^{2}$.

We will now show how to write these equations in terms of a single set
of variables. We introduce a U(2) connection in the following way,
\begin{equation}
{\cal A}_{a} = A_{a}^{i} \sigma_{i} +i \, {\bf
a}_{a} {\bf 1},
\end{equation}
where $e$ is the electric charge, and in a similar fashion a U(2)
electric field,
\begin{equation}
\tilde{\cal E}_{a} = \tilde{E}_{a}^{i} \sigma_{i} + i
 {\bf \tilde{e}}_{a} {\bf 1}.
\end{equation}
That is we are taking the direct product of $U(1)$ and $SU(2)$ to form
a $U(2)$ symmetry. We can similarly introduce a field tensor ${\cal
F}_{ab}$ and a magnetic field.  From these one can recover the
original quantities by taking traces,
\begin{eqnarray}
\tilde{E}^{a}= \tilde{\cal E}^{a} - {\textstyle {1 \over 2}}
Tr(\tilde{\cal E}^{a})\\
\tilde{e}^{a}= {\textstyle {1 \over 2}} Tr(\tilde{\cal E}^{a})
\end{eqnarray}
and so on.

The introduction of these quantities allows us to rewrite the
constraint equations as,
\begin{equation}
D_{a}\tilde{\cal E}^{a}=0
\end{equation}
\begin{equation}
Tr(\tilde{\cal E}^{a} {\cal F}_{ab})=0
\end{equation}
\FL
\begin{eqnarray}
&&{\textstyle {1\over6}} {\rlap{\lower1ex\hbox{$\sim$}}\eta}_{abc}
{\rlap{\lower1ex\hbox{$\sim$}}\eta}_{edf}
Tr(\tilde{\cal E}^{a} \tilde{\cal E}^{b} \tilde{\cal E}^{c})
Tr(\tilde{\cal E}^{e} \tilde{\cal E}^{d} \tilde{\cal B}^{f})+\nonumber\\
&&+{\rlap{\lower1ex\hbox{$\sim$}}\eta}_{abc}
{\rlap{\lower1ex\hbox{$\sim$}}\eta}_{edf} Tr(\tilde{\cal E}^{a} \tilde{\cal
E}^{e}) Tr(\tilde{\cal E}^{b} \tilde{\cal B}^{c}) Tr(\tilde{\cal
E}^{d} \tilde{\cal B}^{f})-\nonumber\\
&&-{\rlap{\lower1ex\hbox{$\sim$}}\eta}_{abc}
{\rlap{\lower1ex\hbox{$\sim$}}\eta}_{edf}
Tr(\tilde{\cal E}^{a} \tilde{\cal E}^{e})
(Tr(\tilde{\cal E}^{c}) Tr(\tilde{\cal E}^{f})+\nonumber\\
&&+Tr(\tilde{\cal B}^{c})
Tr(\tilde{\cal B}^{f})) + \\
&&+ {\textstyle {\Lambda \over 36}}
{\rlap{\lower1ex\hbox{$\sim$}}\eta}_{abc}
{\rlap{\lower1ex\hbox{$\sim$}}\eta}_{edf} Tr(\tilde{\cal E}^{a} \tilde{\cal
E}^{b} \tilde{\cal E}^{c}) Tr(\tilde{\cal E}^{d} \tilde{\cal E}^{e}
\tilde{\cal E}^{f})=0\nonumber
\end{eqnarray}
Notice that we have rescaled the Hamiltonian constraint with a factor
${\rm det}(E)^{2}$ in order to make it polynomial.

It is worthwhile noticing that this is just {\em a rewriting} of the
equations, that is, the theory remains exactly the same. Therefore,
for instance, the constraint algebra and the consistency of the theory
with the reality conditions \cite{AsRoTa} are automatically preserved.

A remarkable fact of this construction is that the ``kinematic''
constraints ---the Gauss Law and the Diffeomorphism constraint--- look
{\em exactly} the same as those of the vacuum theory, only evaluated
for a different group, $U(2)$. This will allow us to exploit several
results found for the vacuum theory for the unified model.

\section{Connection representation \\ and the Chern-Simons form}

We now attempt to quantize the theory in the ``Connection
Representation'', that is, we take a polarization in which
wavefunctions are functionals of the connection $\Psi[{\cal A}]$ and
$\hat{\cal A}_{a} \Psi[{\cal A}] = {\cal A}_{a} \Psi[{\cal A}]$,
$\hat{\tilde{\cal E}}^{a} \Psi[{\cal A}] = {\delta \over \delta {\cal
A}_{a}} \Psi[{\cal A}]$. In the vacuum theory two main results have
been achieved with this representation: a) The result of Jacobson and
Smolin \cite{JaSm} that showed that Wilson loops constructed with
Ashtekar's connection were solutions to the Hamiltonian constraint and
b) The result of Kodama \cite{Ko}, later extended in ref.
\cite{BrGaPunpb}, that showed that the exponential of the Chern-Simons
form constructed with Ashtekar's connection was a solution to all the
constraints with a cosmological constant.

Let us examine these facts for our model. Start by considering
Wilson loops $W_{L}({\cal A}) = Tr( {\rm P\, exp} \oint dy^{a} {\cal
A}_{a}(y))$.  For the vacuum case these quantities solve the
Hamiltonian Constraint.  For our case, however, they fail to be
solutions. This is due to two facts. First of all, since we rescaled
our constraint by the determinant of ${\cal E}$, this quantity should
be nonvanishing for the state of interest. This fact never occurs for
the solutions of the vacuum theory \cite{BrPu}. Even if one considered
a loop with a generic triple intersection, in order to make the
determinant nonvanishing, it is unlikely that finite combinations of
these Wilson loops will be able to solve the constraint. This is
basically due to the fact that the Maxwellian part of the Hamiltonian
constraint has a term ${\tilde {\bf b}}^{2}$ which is purely
multiplicative in this representation and therefore cannot be
cancelled by any of the other terms.

Let us now consider the state,
\begin{equation}
\Psi_{\Lambda}[{\cal A}] = {\rm exp} ( -{ \textstyle {6 \over \Lambda}}
{\rm Tr} \int {\cal A}\wedge d \wedge {\cal A} -{2\over 3} {\cal
A}\wedge{\cal A}\wedge{\cal A})
\end{equation}
It can be readily seen that this state can be decomposed like
$\Psi_{\Lambda}[{\cal A}]= \Psi_{\Lambda}[A] \Psi_{\Lambda}[{\bf a}]$
in terms of the usual Ashtekar variables. It is a remarkable fact that
this state actually manages to solve all the constraints of the theory
with a cosmological constant. This is quite easy to see. The portion
$\Psi_{\Lambda}[A]$ is a solution of the constraints of vacuum gravity
with a cosmological constant
\cite{Ko}. Besides, the portion $\Psi_{\Lambda}[{\bf a}]$ is an
eigenstate of zero eigenvalue for the Hamiltonian of Maxwell theory.
Therefore the product annihilates separately the gravitational and
electromagnetic part of the Hamiltonian constraint. This state will
have important consequences in the loop representation.

What can one say about the physical relevance of this state? The
Chern-Simons form is not a physically relevant state for Maxwell
theory, since it is not normalizable. Evidently there is potential
here for such a problem. Moreover Maxwell theory has a vacuum. Can one
conceive of a state for the combined theory that would bear some
resemblance to the vacuum? These questions will go largely unanswered.
Clearly the Chern-Simons form does not decay fast enough as to be
normalizable in pure Maxwell theory. On the other hand, the notion of
distance that is used to define the vacuum (and to make it fall off
fast enough to assure normalizability) is absent in the combined case,
since the theory is invariant under diffeomorphisms.
This point clearly requires a more careful study before we can reach a
conclusion. Since at the moment there is no candidate for an inner
product for the combined theory we simply cannot say anything about
normalizability.

\section{Loop representation}

The construction of the loop representation for this theory follows
the same steps as those for the vacuum theory so we will only
highlight some points. The reader interested in details of the
construction of loop representations is referred to \cite{RoSm}. The
main difference with the usual case is that the group is $U(2)$
instead of being $SU(2)$. This changes the form of the Mandelstam
identities and therefore the kinematics of the loop representation is
different. As usual we identify,
\begin{equation}
\Psi[L_{1}\circ L_{2}]=\Psi[L_{2}\circ L_{1}].
\end{equation}
but there is no relation between wavefunctions of retraced loops, i.e.
$\Psi[L] \neq \Psi[L^{-1}]$.  In the vacuum theory one also has the
Mandelstam identity, which states,
\begin{equation}
\Psi[L_{1},L_{2}]=\Psi[L_{1}\circ L_{2}] + \Psi[L_{1}\circ L_{2}^{-1}].
\end{equation}
That is, it allows to express any wavefunction of n loops as a
function of n-1 loops. This can be used recursively to reduce all
wavefunctions to functions of only one loop.

For our case the Mandelstam identity now reads,
\FL
\begin{eqnarray}
&&\Psi[L_{1},L_{2},L_{3}] = \Psi[L_{1}\circ L_{2},L_{3}]+
\Psi[L_{2}\circ L_{3},L_{1}]+ \nonumber\\
&&+\Psi[L_{3}\circ L_{1},L_{2}]- \Psi[L_{1}\circ L_{2}
\circ L_{3}] - \Psi[L_{1}\circ
L_{3} \circ L_{2}].
\end{eqnarray}
Where we see that it only allows us to reduce a function of $n$ loops
to a function of $n-1$ and $n-2$ loops. Therefore we cannot reduce all
wavefunctions to functions of only one loop, we need at least two
loops to represent a generic wavefunction. Therefore the kinematics of
the theory is different from the vacuum one, as one may expect from
the fact that the constraint that defines this kinematics, the Gauss
Law, is different in both theories.

An interesting point is that one could proceed in the traditional
fashion (without combining both connections into a $U(2)$ and
construct two loop representations, one for the Ashtekar connection
and another one for the Maxwell one (as was done in \cite{Na} for the
2+1 case). One would then have wavefunctions depending of two loops,
an ``electromagnetic'' and a ``gravitational'' one. Here we also
encounter two loops, but each of them carry information about both
gravitation and electromagnetism. There are other important
differences in the construction of the loop representation in both
cases but we will not discuss them here.

The diffeomorphism constraint still works as a generator of
infinitesimal diffeomorphisms in loop space, and we can represent it
in terms of the area derivative,
\FL
\begin{equation}
\hat{C}(\vec{N}) \Psi[L] = \int d^{3}x N^{a}(x) \oint_{L} dy^{b}
\delta^{3}(x-y) \Delta_{ab}(L_{0}^{y}) \Psi[L].
\end{equation}
The reader is referred to ref. \cite{GaTr} for details of this
expression.  The important point here is that the space of physical
states of the theory will still be represented by functionals of loops
that are invariant under diffeomorphisms, i.e.\ they will be
functionals of the link class of the loop rather than of the loop
itself, exactly as in the vacuum case.

We now turn our attention to the Hamiltonian constraint. Again, it
could be realized in loop space in terms of the area derivative, as
one does for the vacuum theory \cite{Ga}. The calculation is lengthy
and at the moment we do not need the specific form so we will not
exhibit it here. In this case, the constraint is very different from
the vacuum theory, as expected.

One of the main achievements of the loop representation is to make
possible the construction of states that solve {\it all} the
constraints of the theory. We will see that this is the case also for
the model of interest. First of all it is quite simple to see, from
the structure of the Hamiltonian constraint \cite{foot1}, that
wavefunctionals with support on smooth nonintersecting loops are {\em
not} solutions to the constraint. As in the connection representation
this stems from the fact that there is present a purely multiplicative
term in the constraint that fails to annihilate these functionals.
Moreover we should remember that the constraint was rescaled by a
factor that vanishes for loops with less than a triple intersection.

In spite of this we can construct a solution to all the constraints in
the loop representation following the same reasonings that also
allowed to construct solutions in the vacuum case: since the
Chern-Simons form is a solution in the Connection Representation, it
should also be a solution in the loop representation. It turns out
that this transform is known. For our case it turns out to be,
\begin{equation}
\Psi^{CS}_{\Lambda}[L] =  e^{-{\Lambda \over 8} w(L)} J[L](\Lambda).
\end{equation}
Where $w(L)$ is the writhing number of the (framed) loop $L$ and
$J[L](\Lambda)$ is the Jones Polynomial for the loop $L$ in the
variable $\Lambda$; $e$ is the electric charge. This state mimics the
similar one in the vacuum case \cite{BrGaPunpb}.  Notice also that in
order to be a generic state of the theory we need at least two loops.
This presents no difficulty, since the transform of the Chern-Simons
state is well understood for more than one loop \cite{GuMaMi}.

In the loop formulation of gauge theories, it is usual to introduce
charges by opening up the loops. The open path formalism describes
lines of flux with charges at their ends. This has been studied for
the Maxwell theory \cite{GaGr}. It is interesting to note what happens
if one attempts to construct such a formalism for our unified model.
If one opens up one of the loops in question, one not only fails to
satisfy the Gauss Law of the Maxwell theory (which introduces electric
charges) but also one fails to satisfy the Gauss Law of the Ashtekar
formalism. This latter fact only occurs if one couples the theory to
fermions. That is, the loop representation requires that charged
objects should be fermionic.

Another point is that many authors \cite{Ro} have argued that only by
taking into account the quantum properties of the matter that form the
reference frames, physical quantum observables can be defined in
quantum gravity. In that sense, the description of electromagnetic
fields in interaction with gravity could allow to explore an
alternative way for getting physical observables.

We end by mentioning that in the vacuum theory one can perform an
analysis order by order in the Jones Polynomial and retrieve physical
states for the theory without cosmological constant
\cite{BrGaPuessay}. It would be interesting to try to carry out a
similar analysis for the model we are considering.

\section{Conclusions}

We have studied the Ashtekar formulation of the Einstein-Maxwell
theory. We show how one can rewrite the equations in terms of a single
$U(2)$ connection. The kinematic structure of the theory is quite
similar to that of pure General Relativity and allows the
generalization of several results of that case to the combined
Einstein-Maxwell theory. In particular, the loop representation is
quite natural and the Jones Polynomial turns out to be a physical
state of the theory, as happens in pure General Relativity.
Summarizing, we can see that the Ashtekar variables/Loop
Representation approach to the quantization of Gravity can lead to
quite appealing results when one incorporates other interactions in an
unified fashion.

\section{Acknowledgements}

We wish to thank Abhay Ashtekar, Lee Smolin and Charles Torre for
useful discussions. After completing this work we learnt that Ashtekar
and Romano had consider some of these ideas in unpublished work.  This
work was supported in part by grant NSF PHY92-07225 and by research
funds from the University of Utah. J.P.  wishes to thank PEDECIBA and
the Universidad de la Rep\'ublica for hospitality and support in a
visit to Montevideo, where part of this work was accomplished.


\begin{references}
\bibitem{foot1} Although we have not written it in the loop
representation, this conclusion follows
also immediately from the form in the connection representation
\end{references}
\end{document}